\documentclass[11pt]{article}
\usepackage{hyperref}
\pdfoutput=1
\begin{document}
\title{Air entrainment by a plunging jet under \\ intermittent vortex conditions}
\author{Kevin Jin Kim, Kyle Corfman, Kevin Li \&  Ken Kiger\\
\vspace*{6pt} Department of Mechanical Engineering, \\ University of Maryland, College Park, MD 20742, USA}
\maketitle
\begin{abstract}
This fluid dynamic video entry to the 2011 APS-DFD Gallery of Fluid Motion details the transient evolution of the free surface surrounding the impact region of a low-viscosity laminar liquid jet as it enters a quiescent pool. The close-up images depict the destabilization and breakup of the annular air gap and the subsequent entrainment of bubbles into the bulk liquid. 
\end{abstract}
\section*{Description of the Video}
Air entrainment by plunging liquid jets into a pool of the same liquid show a wide range of diverse behavior depending on the viscosity of the fluid, the impact velocity, the interfacial tension, and the disturbance level of the impacting jet (Bin, 1991; Kiger \& Duncan, 2012). The current video focuses on conditions observed when a relatively low viscosity liquid jet impacts as a laminar stream on a quiescent receiving pool of the same liquid. Under these conditions, the entrained subsurface flow can establish a vertically oriented circulation about the axis of the impacting jet. The strength of the circulation and location of its axis varies with time, and when it aligns with the meniscus around the jet, the combined low pressure of the vortex core and the entrainment from the jet is sufficient to generate a symmetric and deep inverted meniscus (sometimes several jet diameters), from which bubbles are entrained from the tip. This mode of air entrainment was first noted by McKeogh \& Ervine (1981), and dubbed the ``intermittent vortex'' regime.

The video presents several sequences showing the formation and destruction of this deep inverted meniscus, acquired at one test condition over a span of several hours. The water jet was formed by a laminar flow nozzle suspended $h$ =  0.165 m above the pool surface, with a flowrate controlled to give an impact velocity of $U_j$ = 3.08 m/s and a jet diameter at impact of $D_j$ = 0.0088 m. In terms of non-dimensional numbers, this gives a Froude number of $Fr = U_j^2/gD_j= 110$, a Weber number of $We =\rho U_j^2 D_j/\sigma = 1160$ and a Reynolds number of $Re = U_j D_j/\nu = 27100$. No perturbations or disturbances are visible on the jet surface, which has the appearance of a smooth glass rod as it enters the pool. The receiving pool was approximately 0.43 $\times$ 0.61 $\times$ 0.43 m in size, with dual overflow weirs on either side to maintain a constant free surface level and minimal surfactant accumulation in the pool.  

When the jet is first started, the meniscus between the liquid jet and the pool curves up and away from the free surface. Beneath the surface, a turbulent jet forms, entraining liquid from the surrounding pool and starting a recirculating flow. As random decaying vorticity in the bulk of the pool is re-ingested into the subsurface jet, the straining field intensifies and aligns the vorticity to form a strengthening circulation in the direction of the jet axis. The low pressure created by the center of this vortex draws down the meniscus, creating an annular sheath of air below the free surface. The vortex contributing to this condition appears to meander and vary in strength, causing the film of air to get shallower or deeper at random intervals. The outer surface of the air film is visibly corrugated with random helical striations, presumably formed by disturbances in the outer flow. 

Once fully formed, the meniscus occasionally destabilizes and ruptures. This process is observed to be initiated by larger fluctuations from the entrained flow or interactions with bubbles in the pool. As the film ruptures, the terminal lip of the sheath where the air is being shed into the pool is contracted into a narrow finger-like structure several millimeters in diameter. The sheath itself then becomes highly asymmetric and degenerates in to a strongly helical structure, gradually reducing in size. It is not known whether this is due to a migration of the circulation or a change in the stability of the film from the change in topology caused by the air entrainment. Air bubbles detached from the sheath are seen to follow a spiraling path along the periphery of the jet. These observations are of interest as they indicate the highly three-dimensional nature governing the air entrainment process and the type of detailed behavior that occurs near the critical entrainment threshold.

\begin{description}
\item [\textsf{Bin A.}] 1993. Gas entrainment by plunging liquid jets. Chemical Engineering
Science 48:3585--3630
\item [\textsf{Kiger K, Duncan J.}] 2012. Air entrainment mechanisms in plunging jets and breaking waves. Annual Reviews of Fluid Mechanics 44: DOI: 10.1146/annurev-fluid-122109-160724
\item [\textsf{McKeogh E, Ervine D.}] 1981. Air entrainment rate and diffusion pattern of plunging
liquid jets. Chemical Engineering Science 36:1161--1172
\end{description}
\end{document}